\documentstyle[11pt,fullpage]{article}
\begin{document}

\begin{flushright}
  hep-th/9702068\\
\end{flushright} \vskip 1.0cm

\renewcommand{\thefootnote}{\fnsymbol{footnote}}
%\bibliography{../onedbits/larefs}
%\bibliographystyle{unsrt}
\def\footnoterule{\kern-3pt \hrule width \hsize \kern2.5pt}
\def\beq{\begin{equation}}
\def\eeq{\end{equation}}
\def\beqn{\begin{eqnarray}}
\def\eeqn{\end{eqnarray}}
\pagestyle{empty}

\begin{center}
{\Large\bf The Size of a Polymer of String-Bits :
A Numerical Investigation\footnote
{Supported in part by
the DoE under grant DE-FG05-86ER-40272 and
by the NSF under grants PHY-93-15811 and PHY-92-18167}
}

\vskip 2cm
\large{Oren Bergman}\footnote{E-mail  address: bergman@string.harvard.edu,
obergman@binah.cc.brandeis.edu}
\vskip 0.5cm
{\it Lyman Laboratory of Physics\\
Harvard University\\
Cambridge, MA 02138}
\centerline{and}
{\it Department of Physics\\
Brandeis University\\
Waltham, MA 02254}
\vskip 1.0cm
\large{Charles B. Thorn}\footnote{E-mail  address: thorn@phys.ufl.edu}
\vskip 0.5cm
{\it Institute for Fundamental Theory\\
Department of Physics, University of Florida\\
Gainesville, FL 32611}

\end{center}

\vspace{1.2cm}
\begin{center}
{\bf ABSTRACT}
\end{center}

\noindent 
In string-bit models, string is described as a polymer of point-like
constituents. We attempt to use 
string-bit ideas to investigate how the
size of string is affected by 
string interactions in a non-perturbative
context.  Lacking adequate methods to deal with the full complications
of bit rearrangement interactions, 
we study instead a simplified analog model with
only ``direct'' potential interactions among the bits. We use the
variational principle in an approximate calculation of
the mean-square size of a polymer as a function
of the number of constituents/bits for various interaction
strengths $g$ in three specific models. 
\vfill
\newpage

\def\balpha{\mbox{\boldmath$\alpha$}}
\def\bgamma{\hbox{\twelvembf\char\number 13}}
\def\bsigma{\hbox{\twelvembf\char\number 27}}
\def\bepsilon{\hbox{\twelvembf\char\number 15}}
\def\sgone{{\cal S}_1{\cal G}}
\def\sgtwo{{\cal S}_2{\cal G}}
\def\ob{\overline}
\def\bx{{\bf x}}
\def\by{{\bf y}}
\def\bz{{\bf z}}
\def\ba{{\bf a}}
\def\bp{{\bf p}}
\def\D{{\cal D}}
\def\R{{\cal R}}
\def\Q{{\cal Q}}
\def\G{{\cal G}}
\def\O{{\cal O}}
\def\N{{\cal N}}
\def\Nlarge{N_c\rightarrow\infty}
\def\Tr{{\rm Tr}}
\newcommand{\ket}[1]{|#1\rangle}
\newcommand{\bra}[1]{\langle#1|}
\newcommand{\firstket}[1]{|#1)}
\newcommand{\firstbra}[1]{(#1|}
\renewcommand{\thesection}{\arabic{section}}
\renewcommand{\thefootnote}{\arabic{footnote}}
\setcounter{footnote}{0}

\pagestyle{plain}
\pagenumbering{arabic}

\setcounter{section}{0}
\section{Introduction}
\renewcommand{\theequation}{\arabic{section}.\arabic{equation}}
\noindent
A long known, perhaps disturbing, fact about string
is its infinite physical extent\cite{susskind70,susskindgrowth}. 
Working in the light-cone gauge, one finds that the 
mean-square transverse size of a string is given by
\begin{equation}
R_\perp^2\equiv\int_0^{P^+/T_0}{T_0d\sigma\over P^+} 
\langle 0|[\bx(\sigma) - \bx(0) ]^2|0\rangle = 
  {d\over\pi T_0}\sum_{n=1}^\infty {1\over n} \longrightarrow \infty \; .
\end{equation}
Any experiment designed to measure this quantity will 
necessarily involve a finite resolution time $\epsilon$.
This means that modes with frequency $>1/\epsilon$ will get
averaged over, and the observed transverse size will
actually be given by:
\begin{equation}
R_\perp^2= 
  2\alpha^\prime d\sum_{n=1}^{\alpha^\prime P^+/\epsilon} 
  {1\over n} \sim 2\alpha^\prime d\ln{\alpha^\prime P^+\over\epsilon}  
  \quad \mbox{for} \quad P^+\gg\epsilon/\alpha^\prime \; ,
\label{cutoffsize}
\end{equation}
where $\alpha^\prime=1/2\pi T_0$ is the Regge slope parameter,
and $d$ denotes the number of transverse dimensions.
As the resolution is improved (smaller $\epsilon$),
the observable transverse extent of string grows logarithmically.
Alternatively, for a fixed $\epsilon$ the transverse size 
grows with increasing longitudinal momentum. Since the growth
is logarithmic it will barely be noticeable unless the string
experiences very high longitudinal boosts, such as when it falls
into a black hole
\cite{susskindbh1,susskindbh2}. 

This peculiar property of perturbative strings, known as
branching diffusion, is well known from the days
of dual resonance models \cite{susskind70}. The form of the 
scattering amplitude at high energy and 
fixed momentum transfer
(Regge regime),
\begin{equation}
 A \sim s^{1-t} = e^{(1-t)\ln s} \; ,
\end{equation}
implies a transverse target size $R^2_\perp\sim\ln s$. 
Consider a process in which one of the strings carries most
of the longitudinal momentum, $P^+_1\gg P^+_2$. Then in 
a frame where the transverse momenta are equal and opposite,
${\bf p}_1=-{\bf p}_2\equiv{\bf p}$, the c.o.m. (energy)$^2$ is
given by
\begin{equation}
 s = 2(P_1^+ + P_2^+)(P_1^- + P_2^-) 
\sim |\bp|^2{P^+_1\over P^+_2} = 2P^+_1E_2 \; .
\end{equation}
The transverse size is therefore given by
\begin{equation}
 R_\perp^2 \sim \ln 2P^+_1E_2 \sim \ln{P^+_1\over P^+_2} \; .
\end{equation}
Comparing with (\ref{cutoffsize}), we see that the resolution time
$\epsilon$ is given by the longitudinal momentum of the small string,
$\epsilon = \alpha^\prime P^+_2$. Since the energy of this ``probe'' string
is given by ${\bf p}^2/(2P^+_2)$, the transverse size of the 
``target'' string will appear
to grow as the energy of the probe string is increased.
In string theory $P^+$ is a continuous variable, so in 
particular the probe string can have $P^+_2 = 0$, in which
case the target string's size will be infinite.

Consider discretizing a piece of string into $M$ segments,
such that each segment carries a
longitudinal momentum $\Delta P^+$. 
The total longitudinal momentum is then $M\Delta P^+$.
This means that the smallest
probe string has $P^+_2 = \Delta P^+$, and that given this 
probe, the target string grows logarithmically with $P^+_1\equiv P^+$. 
String-bit models give precisely such a dicretization, where
$\Delta P^+$ is given by the bit mass $m$.
This provides a 
{\em physical} cutoff, resulting in a transverse size given by
\begin{equation}
 R_\perp^2 \sim {d\over\pi T_0}\ln{P^+\over m} 
   = {d\over\pi T_0}\ln M\; ,
\end{equation}
where $M$ is the number of bits in the target string. Comparison
with the previous discussion shows that we can identify 
$\epsilon=m\alpha^\prime$.
As the transverse size is a measure of the transverse volume occupied
by the string, the transverse bit number density of string is given by
\begin{equation}
 \rho_\perp = {M\over R_\perp^d} 
  \sim \left({\pi T_0\over d}\right)^{d/2}{M\over (\ln{M})^{d/2}}\; .
\end{equation}
String interactions arise in string-bit models through the
rearrangement interaction of string-bits. In perturbation
theory a longitudinal dimension ($x^-$) emerges dynamically.
Thus the effective interaction of string is measured
by the product of the string coupling and the bit number
density in this higher dimension space. The longitudinal
size of noninteracting string is given by
\begin{equation}
R_L^2\equiv\int_0^{P^+/T_0}{T_0d\sigma\over P^+} 
 \langle 0|\big[x^-(\sigma) - x^-(0)\big]^2|0\rangle 
  \sim {d\over3P^{+2}}\sum_{n=1}^{\alpha^\prime P^+/\epsilon} n 
  \sim {d\over6}\left({\alpha^\prime\over\epsilon}\right)^2 \; ,
\end{equation} from which we see that $R_L$ doesn't grow
with $P^+$  at fixed $\epsilon$, but is nonetheless very large 
$O(\alpha^\prime/\epsilon=1/m)$. Thus the effective bit number
density relevant to weakly interacting string is
\begin{equation}
\rho_{eff}={\rho_\perp\over R_L}
        \sim \sqrt{6\over d}{P^+\over(2d\alpha^{\prime}
\ln(\alpha^\prime P^+/\epsilon))^{d/2}}.
\end{equation}
For {\it fixed} $\epsilon$ ($m$ in string
bit models) this density grows essentially
linearly with increasing $P^+$. It will eventually become
comparable to $1/g^2$, where $g$ is the string coupling constant.
At this point interactions become important and
perturbation theory must break down. Since the above formulas 
rely on a perturbative picture, they cannot be correct at 
arbitrarily high $P^+$. At such energies the integrity
of string is lost and a more appropriate description
would presumably be in terms of a fluid of string-bits. 

It has been suggested that
one effect of string interactions is to spread string out,
in essence to push bits away from each other \cite{susskindbh2}. 
In fact it has been 
conjectured that the transverse size should grow just rapidly
enough to yield 
a constant transverse bit number density of $m^d_{Planck}$ as 
$P^+\rightarrow\infty$.
This would be necessary if string theory is to account
for the Beckenstein-Hawking black-hole entropy \cite{Hawk,Beck}, and 
is to provide a realization of 't~Hooft's holographic principle
\cite{thoofthol}. This limiting transverse
density translates to a lower bound on the
transverse size:
\begin{equation}
 R_\perp^2 \buildrel>\over\sim {M^{2/d}\over m_{Planck}^2} \; .
\label{lineargrowth}
\end{equation}
For $d=2$ ($D=4$ space-time) this lower bound is simply
linear in bit number.

The only indication in perturbative string theory that the
net effect of the interactions is repulsive is the
underlying supersymmetry, which guarantees 
a positive semidefinite energy. 
In the context of 
supersymmetric string-bit models, since bits must be held together by
attractive interactions, it follows that residual interactions
between composites of bits ({\it i.e.} string) should on the average be 
repulsive. 
It is clear that further insight into this issue
requires information beyond perturbation theory.
We propose to study string size in the context
of $d=2$ string-bit models, which are particular composite
formulations of $D=4$ string theory, that (conveniently) incorporate
perturbative string interactions in the larger setting
of microscopic bit interactions.

The complete dynamics of string-bit models are 
extremely complicated.
The interactions among bits include, among other complications,
rearrangements of the bits among different polymers as well as
on a single polymer. We have not yet developed methods to
handle these complications, so as a preliminary exploration,
we consider a simple many-body quantum mechanical system
of interacting polymers, motivated by the general principles
of string-bit models. This simplified model contains only
``direct'' (potential energy) interactions, i.e. no exchange
effects. Admittedly, this system is 
too simple to capture the quantitative details of non-perturbative
string growth,
but we can hope that it (or a slight improvement of it) can capture
some of the qualitative physics. As we shall see, the growth predicted
by this model is actually too rapid: quadratic in bit number rather
than linear. It is too early to tell whether this ``overshoot'' 
signals a problem with the bit model itself, or it simply indicates 
the crudeness with which our analog system imitates a string-bit
model.

The rest of the paper is organized as follows.
In section 2 we discuss the size of generic extended 
many-body bound states in 
such
potential models, and show that the condition
that these objects follow a relativistic dispersion law
($P^-=({\bf p}^2+{\cal M}^2)/2P^+$) implies an upper bound
on the growth rate. In section 3 we rederive the logarithmic
transverse growth of free string as the growth of a 
polymer of string-bits in the $N_c\rightarrow\infty$ limit. 
In section 4 we consider
the effect of finite $g\sim1/N_c$, and propose quantum mechanical
models which mimic it. In section 5 we 
use a variational approach to compute, in three specific models, 
the mean square size of a polymer and the bond length as
functions of the number of bits. Section 6 is devoted to a discussion
of the results and future directions.

\section{Size of Extended Bound States}
\setcounter{equation}{0}
\noindent
Consider a many particle system in $d$ space dimensions, with dynamics 
governed by the Hamiltonian
\begin{equation}
 h = \sum_{k=1}^M{\bp_k^2\over 2m} + V(\bx_1,\ldots,\bx_M)\; .
\end{equation}
Assuming that this system admits extended bound states of many particles
(e.g. chains), what is their rate of growth with the number of constituents?
We will discover that for these bound states to become relativistic objects
in $d+2$ space-time dimensions in
the continuum limit, the mean-square size can grow at most linearly. 

The mean-square distance between the $(k+l)$'th 
constituent and the $l$'th constituent (``$k$-separated'' constituents)
on the object in its ground state is given by:
\begin{equation}
  R_k^2 = \langle 0|(\bx_{k+l} - \bx_l)^2|0\rangle = {4\over M}
  \sum_{n=1}^{M-1}\sin^2{\pi nk\over M} 
    \bra{0}\hat{\bx}_n\cdot\hat{\bx}_{M-n}\ket{0}\; ,
  \end{equation}
where $\hat{\bx}_n$ are the Fourier components of $\bx_k$:
\begin{equation}
 \hat{\bx}_n = {1\over \sqrt{M}} \sum_{k=1}^M {\bf x}_k e^{2\pi ink/M}  \; .
\label{fourier}
\end{equation}
We define the  mean-square size as the average of the mean-square distances
\begin{eqnarray}
 R^2 &\equiv& {1\over M} \sum_{k=1}^M R_k^2 \nonumber\\
 &=& {2\over M} \sum_{n=1}^{M-1}\bra{0}\hat{\bx}_n\cdot\hat{\bx}_{M-n}\ket{0} \; .
\label{genericsize}
\end{eqnarray}
Note that the variable conjugate to $\hat{\bx}_n$ is $\hat{\bp}_{M-n}$, so
\begin{equation}
 -i\sum_i \bra{0}[\hat{x}^i_n,\hat{p}^i_{M-n}]\ket{0} = d\; .
\end{equation}
Using $\hat{p}^i_n = -im[\hat{x}^i_n,h]$ and inserting complete sets of states we get
\begin{equation}
 d = m\sum_{\lambda}(E_\lambda - E_0) \Big[ 
       \bra{0}\hat{\bx}_n\ket{\lambda}\cdot\bra{\lambda}\hat{\bx}_{M-n}\ket{0}
    + \bra{0}\hat{\bx}_{M-n}\ket{\lambda}\cdot\bra{\lambda}\hat{\bx}_n\ket{0}\Big] .
\end{equation}
Since each term in the sum is positive definite, 
and $E_{\lambda_1}\geq E_{\lambda_2}$ for $\lambda_1>\lambda_2$, we get the
following inequality:
\begin{equation}
 d \geq 2m(E_1 - E_0)(\bra{0}\hat{\bx}_n\cdot\hat{\bx}_{M-n}\ket{0}
 -\bra{0}\hat{\bx}_n\ket{0}\bra{0}\hat{\bx}_{M-n}\ket{0}) \; .
\end{equation}
If the Hamiltonian possesses at least a cyclic symmetry with
respect to permutations of the particles, $\bra{0}{\bf x}_k\ket{0}$
will be independent of particle label $k$, 
in which case (\ref{fourier}) shows that $\bra{0}{\bf\hat x}_n\ket{0}=0$ for $n\neq 0$, and the second term vanishes.
We then find
\begin{equation}
 d \geq 2m(E_1 - E_0)\bra{0}
\hat{\bx}_n\cdot\hat{\bx}_{M-n}\ket{0} \; .
\end{equation}
{}From eq.~(\ref{genericsize}) we therefore get an 
upper bound on the size:
\begin{equation}
 R^2 \leq {d\over m}{M-1\over M}{1\over E_1 - E_0} \sim {d\over m(E_1 - E_0)} \; .
\end{equation}
The only question remaining is how the gap $E_1 - E_0$ scales with $M$.
If the extended object is to describe a discretization of a {\em relativistic} object
in the light-cone frame (e.g. chain of bits $\rightarrow$ light-cone string),
then the gap should scale as $1/mM$. This is because in the continuum limit
$mM\rightarrow P^+$, so 
\begin{equation}
 \Delta E = \Delta P^- = {{\cal M}^2\over 2P^+} \sim {{\cal M}^2\over 2mM} \; ,
\end{equation}
where ${\cal M}$ is the rest mass of the lightest massive 
particle.
Consequently the bound becomes:
\begin{equation}
 R^2 {\ \lower-1.2pt\vbox{\hbox{\rlap{$<$}\lower5pt\vbox{\hbox{$\sim$}}}}\ } 
 {2d\over{\cal M}^2}M= {2d\over m{\cal M}^2} P^+ \; .
\end{equation}
Note that we have not used a harmonic (or any specific) interaction 
(nearest-neighbor
or not) to derive this bound. It is simply a consequence of the fact that the 
potential depends only on the positions and not on the momenta.

A related upper bound on the growth of relativistic strings was
derived in a completely different context by Susskind 
\cite{susskindhol}. This bound is due to causality
in the presence of a background
black hole geometry. The information carried by a spreading string on the 
stretched horizon is limited in its speed of propagation
by the requirement that it lie inside the light cone.
The bound on the spread of information is found to be
\begin{equation}
 R_\perp^2  {\ \lower-1.2pt\vbox{\hbox{\rlap{$<$}\lower5pt\vbox{\hbox{$\sim$}}}}\ } 
  e^{t/4M_{BH}G} \; ,
\end{equation}
where $t$ is time as measured by an external (Schwarzschild) observer,
and $G$ is Newton's constant.
Since the longitudinal momentum of the string in this situation 
grows as
\begin{equation}
 P^+ \sim e^{t/4M_{BH}G} \; ,
\end{equation}
it follows that the maximum rate of growth of the mean-square size with
longitudinal momentum is linear. 
Here too, the bound is independent of the number of dimensions, and 
happens to agree with the conjectured growth rate in four space-time dimensions
($d=2$).

In the next section we shall see that a bare polymer of string-bits,
i.e. a free string, grows logarithmically, and therefore satisfies 
the bound (2.11) for any $d$. Moreover, for $d\geq2$ this upper
bound is compatible with the lower bound in eq.~(\ref{lineargrowth}). 
Note that for $d=1$, i.e. three space-time
dimensions, the two bounds are incompatible.
{}From the point of view of extended bound states 
the $d=1$ growth pattern is trivially
understood, since the size of a one dimensional chain is simply its length.
{}From the above argument it follows 
that the continuum limit cannot correspond to 
a relativistic string. The black-hole causality argument is
also problematic for $d=1$.

\section{Size of a Bare Polymer of String-Bits}
\setcounter{equation}{0}
\noindent
In the $N_c\rightarrow\infty$ limit, the energy eigenstates 
of string-bit models 
\cite{thornmosc,bergmantbits,bergmanevidence,thorn96} 
are non-interacting (bare) multi-polymer states.
A single bare polymer containing $M$ bits 
has, for $M\to\infty$, physical properties identical to
a relativistic string in light-cone gauge.
The dynamics of a bare polymer is governed by a Hamiltonian with
only ``nearest-neighbor'' interactions: 
\begin{equation}
h = {1\over 2m}\sum_{k=1}^M \Big[ {\bf p}_k^2 + V({\bf x}_{k+1}-{\bf x}_k) \Big] \; ,
\label{manybodyprob}
\end{equation}
where $V(\bx)$ is an attractive (and binding) potential.
For the special case of harmonic interactions this becomes
\begin{eqnarray}
 h &=& {1\over 2m}\sum_{k=1}^M
\Big[\bp_k^2 + T_0^2(\bx_{k+1} - \bx_k)^2\Big]\nonumber\\
 & = & {1\over 2m}\sum_{n=1}^{M-1}\Big[\hat{\bp}_n\cdot\hat{\bp}_{M-n} 
            + \omega_n^2\hat{\bx}_n \cdot\hat{\bx}_{M-n}\Big] \; ,
\label{barechain}
\end{eqnarray}
and the Hamiltonian 
can be diagonalized in terms of the creation and annihilation
operators given by
\begin{eqnarray}
 \ba_n &=& {1\over\sqrt{2\omega_n}}\left(\hat{\bp}_n 
 - i\omega_n\hat{\bx}_n\right) \nonumber \\
 \ba^\dagger_n &=& 
      {1\over\sqrt{2\omega_n}}\left(\hat{\bp}_{M-n} + i\omega_n\hat{\bx}_{M-n}\right) \; .
\end{eqnarray}
Using these in the equations of the previous section gives
\begin{equation}
 R_k^2 = \langle 0|(\bx_{k+l} - \bx_l)^2|0\rangle = {2d\over M}
  \sum_{n=1}^{M-1}{\sin^2\pi nk/M\over\omega_n} \; ,
\label{size_k}
\end{equation}
and
\begin{equation}
 R^2 = {1\over M}\sum_{k=1}^M R_k^2 = 
  {d\over M}\sum_{n=1}^{M-1} {1\over\omega_n} \; ,
\label{size}
\end{equation}
for the mean-square distance between $k$-separated bits and
the mean-square size of the chain, respectively.
The above formulas hold for {\em any} harmonic system with
normal mode frequencies $\omega_n$. For the nearest-neighbor
harmonic system of a bare chain (\ref{barechain}): 
$\omega_n = 2T_0\sin n\pi/M$, so
\begin{equation}
 R^2 \sim {d\over\pi T_0}\ln M \; ,
\label{baresize}
\end{equation}
in agreement with the free string result.

For a general potential, an exact solution is impossible and we turn
to the approximation methods developed in \cite{thornrpa}.
Technically it is easier to deal in the general case with an open
polymer with the Hamiltonian
\begin{equation}
h_{\rm open}={1\over 2m} \sum_{i=1}^M {\bf p_i}^2
 + {1\over 2m}\sum_{i=1}^{M-1}V({\bf x}_{i+1}-{\bf x}_i).
\label{openpolymer}
\end{equation}
Since the size of open polymers with a large number of bits is easily
related to the size of closed polymers with the same large number of
bits, the conclusions will be similar.
Because of the nearest-neighbor interaction pattern,
a particularly neat way of separating the center of mass motion is
to define internal coordinates ${\bf y}_k={\bf x}_{k+1}-{\bf x}_k$, with
corresponding conjugate momenta ${\bf q}_k$, in which case the 
Hamiltonian for internal dynamics becomes
\begin{equation}
h^\prime_{{\rm open}}\equiv{1\over m}\left[
\sum_{r=2}^M {\bf q}_r^2-\sum_{r=2}^{M-1}{\bf q}_r\cdot{\bf q}_{r+1}
+ {1\over 2} \sum_{r=2}^M V({\bf y}_r)\right].\label{openpolymerham}
\end{equation}
The correlation functions $\bra{G}T[y_k(t)y_l(0)]\ket{G}$ contain
information about the excitation spectrum of the model as well as
information about the system's size. Spectral information is inferred
by a Fourier analysis of the time dependence, while size information
is given by the limit $t\to0$. Define the Fourier components of
this correlator by
\beqn
\bra{G} T[y_r^i (t) y_s^j(0)]\ket{G} &\equiv& - \int {d\omega\over
2\pi i} {\cal E}_{rs}^{ij} (\omega) e^{-i\omega t} \; .
\eeqn
As discussed in \cite{thornrpa}, it is convenient to introduce
a certain ``irreducible'' two point correlator $\Pi^{ij}_{rs}(\omega)$
which facilitates the analysis of the low energy spectrum in the
limit of large bit number. The irreducibility is with respect to
a graphical description of time dependent perturbation theory, taking
the term $\sum q_rq_{r+1}$ as the perturbation. For example the lowest
order contribution to $\Pi$ is just ${\cal E}$ for the system with
this perturbation set to zero. In general $\Pi$ is the sum of all
connected 
graphs contributing to ${\cal E}$ which can not be disconnected
by cutting only one line (which graphically represents the interaction).
The all orders relation of ${\cal E}$ to $\Pi$ is then, assuming rotation
invariance:  
\beq
{\cal E}_{rs}^{ij}(\omega)=\delta_{ij}
\left[\Pi^{-1}-{m^2\omega^2\over2}\left(1-{2\over k^2}
 \right)\right]^{-1}_{rs}\; .
\eeq
Here $(k^2)_{rs}=2\delta_{rs}-\delta_{rs+1}-\delta_{rs-1}$.
The eigenvalues of the matrix $(k^2)_{rs}$ are $4\sin^2(n\pi/2M)$, some of which
scale as $1/M^2$ at large $M$. Thus if $\Pi_{rs}$ is proportional
to $\delta_{rs}$ and non-vanishing at $\omega=0$, ${\cal E}$ will display 
poles at
$\omega=O(1/M)$, as required for a stringy continuum limit. This is
certainly the case for a harmonic potential for which $\Pi_{rs}$ is exactly
given by $\Pi_{rs}=2\delta_{rs}/(m^2\omega^2-2T_0^2)$. Of course for
a general potential one can't solve exactly for $\Pi$, but one can
argue that it will generically have this low 
frequency behavior for $M\to\infty$
for a wide class of potentials.

On the other hand the integral of ${\cal E}$ over all $\omega$ gives
the limit $t=0$, which contains size information. Doing the integral
by contours gives a contribution to the size from the above
mentioned low energy poles
similar to the purely harmonic nearest-neighbor case, 
i.e. a contribution scaling as
$\ln M$. But there are of course other contributions which could
conceivably alter this behavior. For the  
nearest-neighbor harmonic
case this doesn't
happen; the integral can be exactly performed and the conclusion
of logarithmic growth is unaltered. We suspect this conclusion is
more general, but our lack of detailed knowledge of the behavior of
$\Pi$ at high frequency leaves the matter open to future study.

\section{Bits With Elbows -- A Toy Model}
\setcounter{equation}{0}
\noindent
In order to carry the discussion of size beyond bare polymers
(free string) we must deal with the complete string-bit
Hamiltonian at finite $N_c$. The simplest (bosonic) string
bit model \cite{thornmosc} is described by the Hamiltonian:
\begin{equation}
H=\int d{\bf x}{1\over2m}\Tr\nabla\phi^\dagger\cdot\nabla\phi 
    - {1\over 4mN_c}\int d{\bf x}d{\bf y}V({\bf x}-{\bf y})
  :\Tr[\phi^\dagger({\bf x}),\phi({\bf x})]
    [\phi^\dagger({\bf y}),\phi({\bf y})] : \; ,
\label{secondham}
\end{equation}
where $ \phi^{\dagger \beta}_{\alpha}({\bf x})$
creates a bit at the position ${\bf x}$ with color $(\alpha,\beta)$, and
lower (upper) color indices transform in the $N_c$ (${\bar N}_c$)
representations of $SU(N_c)$.

The variational principle provides a natural approach to the
problem of discovering how string interactions ($1/N_c$ corrections) 
affect the size of string. Since string-bit dynamics is standard
many body non-relativistic quantum mechanics, exact energy
eigenstates are given by states for which the functional
$$E[\psi]\equiv {\bra{\psi}H\ket{\psi}\over\langle{\psi}|\psi\rangle}$$
is stationary with respect to arbitrary infinitesimal changes
$\delta\ket{\psi}$. Furthermore the lowest energy eigenstate is the
absolute minimum of this functional. Since the Hamiltonian is
a color singlet and commutes with 
bit number, one loses no generality in restricting $\ket{\psi}$
to an irreducible representation of the color group and to contain
a fixed number $M$ of string-bits.
Focusing on the
color singlet sector, a general trial state $\ket{\psi}$ is a linear
combination of multi-polymer states of the form 
(assuming $M\geq k_{n-1}\geq\cdots\geq k_2\geq k_1\geq 1$)
\begin{eqnarray}
 \ket{\psi}^{(k_1,\cdots,k_{n-1})}_M 
          =  N_c^{-{M/2}}  \! \! \int \! d\bx_1\cdots d\bx_M 
         {\rm Tr}[\phi^\dagger(\bx_1)\cdots
             \phi^\dagger(\bx_{k_1})] \,
            {\rm Tr}[\phi^\dagger(\bx_{k_1+1})\cdots
             \phi^\dagger(\bx_{k_2})] \cdots \nonumber \\
     \cdots{\rm Tr}[\phi^\dagger(\bx_{k_{n-1}+1})\cdots
             \phi^\dagger(\bx_{M})]\ket{0} 
\psi^{(k_1,\cdots,k_{n-1})}_M(\bx_1,\ldots,\bx_M) \; . \qquad
\end{eqnarray}
At $N_c=\infty$ the Hamiltonian acts independently on each trace,
showing that in this limit the above state 
describes $n$ noninteracting bare polymers, whose energy eigenstates
are described by the chain dynamics explained in the previous section.

Of course, making unrestricted variations is tantamount to 
solving the problem exactly, a presumably intractable
task. To proceed in an approximate way, 
we seek a minimum of $E[\psi]$ within a restricted class of states.
In this paper we restrict $\ket{\psi}$ to be a single
bare polymer 
$$
  \ket{\psi_M}^{(1)} = N_c^{-M/2}\int d\bx_1\cdots d\bx_M
             {\rm Tr}[\phi^\dagger(\bx_1)\cdots
             \phi^\dagger(\bx_M)]\ket{0}\psi_M(\bx_1,..,\bx_M) \; ,
$$
with $\psi_M$ restricted to a class of functions for which
integrals can be readily performed. The approximate
Schr\"odinger dynamics that follows by using this trial
in the variational principle assumes the form
\begin{eqnarray}
\bra{0}{\rm Tr}[\phi(\bx_M)\cdots
             \phi(\bx_1)](H-E[\psi])\ket{\psi_M}^{(1)}=0
\end{eqnarray}
The matrix element of $H$ may be reduced by letting $H$ act
to the right. 
Let $H^\prime$ be the interaction term in (\ref{secondham}).
Its action on the one-polymer state is given by
\goodbreak
\begin{eqnarray}
H^\prime
\ket{\psi_M}^{(1)}
&=& - {N_c^{-M/2}\over 4mN_c}\int d\by_1\cdots d\by_M\sum_{i<j}
[V(\by_i-\by_j)+V(\by_j-\by_i)]  \nonumber\\
& \bigg\{ & {\rm Tr}[\phi^\dagger(\by_i)\cdots
             \phi^\dagger(\by_{j-1})]{\rm Tr}[\phi^\dagger(\by_j)\cdots
             \phi^\dagger(\by_{i-1})]\nonumber\\
& + & {\rm Tr}[\phi^\dagger(\by_{i+1})\cdots
             \phi^\dagger(\by_{j})]{\rm Tr}[\phi^\dagger(\by_{j+1})\cdots
             \phi^\dagger(\by_{i})] \nonumber \\[5pt]
&  - &  {\rm Tr}[\phi^\dagger(\by_{i})\cdots
             \phi^\dagger(\by_{j})]{\rm Tr}[\phi^\dagger(\by_{j+1})\cdots
             \phi^\dagger(\by_{i-1})]\nonumber\\
& - & {\rm Tr}[\phi^\dagger(\by_{i+1})\cdots
             \phi^\dagger(\by_{j-1})]{\rm Tr}[\phi^\dagger(\by_{j})\cdots
             \phi^\dagger(\by_{i})]
              \; \bigg\} \ket{0} \, \psi_M(\by_1,..,\by_M) \; .
\label{interactions}
\end{eqnarray}
When one of the traces in (\ref{interactions}) is empty, 
that trace simply provides a factor
of ${\rm Tr}\ I=N_c$ which cancels the $1/N_c$ out front thus
providing  a term that survives the $N_c\to\infty$ limit. Simple
inspection shows that this only happens for the last term in 
braces when $j=i+1$ and for the third term in braces when $i=1$ and $j=M$.
We combine these special terms with the terms coming from
the matrix element of the kinetic term of $H$ to give the
first quantized Hamiltonian $\hat h$ of a bare polymer,
defined by:
\begin{equation}
\bra{0}{\rm Tr}[\phi(\bx_M)\cdots\phi(\bx_1)]|{\hat h}\psi_M\rangle^{(1)} 
                   \equiv
  \lim_{N_c\rightarrow\infty} 
\bra{0}{\rm Tr}[\phi(\bx_M)
                \cdots\phi(\bx_1)]H \ket{\psi_M}^{(1)} \; ,
\end{equation}
and therefore given by
\begin{equation}
 {\hat h} = {1\over 2m}\sum_{k=1}^M  \bigg[\bp_k^2  +   
              V(\bx_{k+1}-\bx_k)\bigg]\; ,
\end{equation}
where we have, without loss of generality, taken the potential to be even
$V({\bf x})=V(-{\bf x})$.

All the remaining terms in (\ref{interactions}) vanish in
the large $N_c$ limit. At finite $N_c$, they introduce several new
physical features into the dynamics. First of all, they
describe interactions between non-nearest-neighbors on
the bare polymer. At the same time, they allow non-cyclic
bit rearrangement
on the polymer. Finally, there are contributions of both
an attractive and a repulsive character. Because the exact
Hamiltonian is positive, we can expect that on average
the non-nearest-neighbor interactions are repulsive.
We can roughly confirm this by counting the number of repulsive 
and attractive contributions to
the matrix element of $H^\prime$ at leading order in $1/N_c$.
For example, the contribution of the first term in
(\ref{interactions}) is given by
$$
 N_c^{-M-1}\bra{0}\Tr [\phi(\bx_M)\cdot\cdot \phi(\bx_1)] 
 \Tr [\phi^\dagger(\by_{i})\cdot\cdot
             \phi^\dagger(\by_{j-1})]
 \Tr[\phi^\dagger(\by_{j})\cdot\cdot
             \phi^\dagger(\by_{i-1})] \ket{0} \, .
$$
To evaluate the above matrix element we simply contract the
annihilation and creation operators. The leading
order contribution corresponds to contractions which 
preserve the cyclic ordering in the traces. 
There are therefore $M(j-i)(M-j+i)$ such terms
in the above matrix element,
each giving an additional factor of $N_c^{M-1}$, for
a total of $N_c^{-2}$.
{}From eq.~(\ref{interactions}) it then follows that there
are $2M(j-i)(M-j+i)$ repulsive terms and 
$M(j-i+1)(M-j+i-1) + M(j-i-1)(M-j+i+1) = 2M(j-i)(M-j+i) - 2M$
attractive terms; an excess of repulsive over attractive
of $2M$. 
(Note that the factor of $M$ in these countings is
common to all terms in the Hamiltonian matrix element and does
not represent an undue enhancement.) 

Handling bit rearrangement effects in the context of a variational
calculation is daunting if not intractable. We defer such a
direct attack and instead, in a first attempt,  replace our string-bit system
with an effective analog system in which rearrangement is not allowed.
This strategy is somewhat reminiscent of the Hartree as opposed
to Hartree-Fock approximation to many electron atoms. We wish to
suppress rearrangement of bits altogether and simulate $1/N_c$
effects by adding a simple non-nearest-neighbor potential
term to ${\hat h}$:  
\begin{equation}
   h = {1\over 2m}\sum_{k=1}^M\Big[\bp_k^2 + V(\bx_{k+1} - \bx_k)\Big]
   - {\xi\over N_c^2}\sum_{k\neq l} V(\bx_k - \bx_l).
\end{equation}
The last term is supposed to represent, in an average way, all
of the terms in (\ref{interactions}) with nonempty traces. 
We have taken it to be
repulsive because although there are a large number of terms with
both signs, there is a slight preponderance of repulsive ones. We
have used the same potential function for the nearest-neighbor
attraction as the non-nearest-neighbor repulsions, as both
effects originate from the same quartic term in the Hamiltonian.
But we have introduced the parameter $\xi$ to absorb the 
renormalization effects of the averaging procedure.

Furthermore, for the sake of numerical study we would like to relax the 
condition that the non-nearest-neighbor interaction be tied to
the nearest-neighbor interaction. 
Thus we shall study the ``bits with elbows'' system given by
\begin{equation}
   h = {1\over 2m}\sum_{k=1}^M\Big[\bp_k^2 + V(\bx_{k+1} - \bx_k)\Big]
   + g^2\sum_{k\neq l} U(\bx_k - \bx_l) \; ,
\end{equation}
where $U(\bx)$ is an independent  repulsive potential, and
$g^2$ represents $1/N_c^2$ together with all renormalization effects.
This will allow us in particular to use
a long-range (harmonic) nearest-neighbor potential 
in conjunction with a short-range non-nearest-neighbor potential.

Consider first the harmonic chain of the previous section, $V(\bx)\sim \bx^2$.
For small $g$, one might be tempted to treat the 
non-nearest-neighbor terms as a perturbation. The resulting
first order correction to the size of the chain is found to be
\begin{equation}
 \Delta R^2 \sim g^2 M^3 [\ln M]^\alpha \; ,
\end{equation}
where $\alpha$ depends on the precise form of $U(\bx)$. Comparing
with the zeroth order result (\ref{baresize}), we see that
perturbation theory breaks down quickly with increasing number
of bits. To treat a large number of bits we will instead use a variational
approach.

\section{Variational Approach and Numerical Results}
\setcounter{equation}{0}
\noindent Although we have used the variational principle to motivate our analog model,
the model itself cannot be solved without approximation. Thus
in this section we shall use the variational method to learn about
the size properties of the analog system. If we make simple choices for 
$V$ and $U$, the integrals involved in computing $E[\psi]$ can
be readily done once we restrict the trial wave functions to be gaussians.
A convenient way to parameterize such trial functions is to let them
be the ground state wave functions of various
many body systems with arbitrary harmonic forces. The
ground state of such a system is defined by\footnote{First-quantized states are
denoted $\firstket{\cdots}$, to distinguish them from second-quantized (Fock)
states $\ket{\cdots}$.}
\begin{equation}
 A_n\firstket{\phi} = 0 \; ,
\end{equation}
where 
\begin{equation}
 A_n = {1\over \sqrt{2\omega_n}}(\hat{\bp}_n - i\omega_n\hat{\bx}_n) \; .
 \end{equation}
We take the normal mode frequencies  $\omega_n$ of the  
trial harmonic system 
as the variational parameters, and minimize the expectation value
of the energy 
\begin{equation}
 E(\omega_n) = {\firstbra{\phi} h \firstket{\phi}\over(\phi|\phi)}
\end{equation}
with respect to them.
The size of the chain will then be given by eq.~(\ref{size}).
In addition we will want to keep track of the bond length,
given by eq.~(\ref{size_k}) with $k=1$:
\begin{equation}
 R_1^2 = {2d\over M}\sum_{n=1}^{M-1} {\sin^2{\pi n/M} \over \omega_n} \; .
\end{equation}
This will determine the string scale in the stringy
continuum limit ($M\to\infty$).
Three specific models are considered:
\begin{enumerate}
 \item Harmonic chain with $\delta$-function elbows:
 \begin{equation}
   V(\bx) = T_0^2 \bx^2 \quad , \quad U(\bx) = \delta^{(d)}(\bx) \; .
 \end{equation}
 Setting $m=1$ and absorbing numerical factors into $g^2$, 
 the expectation value for the energy is 
 \begin{equation}
   E_{h\delta}(\omega_n)= {1\over 2}\sum_{n=1}^{M-1} \omega_n
      + {1\over 2} T_0^2 M R_1^2 
      + g^2 M \sum_{k=1}^{M} {1\over R_k^d} \, .
 \end{equation}
 \item Harmonic chain with gaussian elbows:
 \begin{equation}
   V(\bx) = T_0^2 \bx^2 \quad , \quad U(\bx) = e^{-\bx^2/a^2} \; ,
 \end{equation}
 \begin{equation}
    E_{hg}(\omega_n)= {1\over 2}\sum_{n=1}^{M-1} \omega_n
      + {1\over 2} T_0^2 M R_1^2 
      +  g^2 M \sum_{k=1}^{M} \left[{a^2d\over 2} + R_k^2\right]^{-d/2}\, .
 \end{equation}
 \item Gaussian chain with gaussian elbows:
 \begin{equation} 
  V(\bx) = - \lambda e^{-\bx^2/a^2} 
                 \quad , \quad U(\bx) = e^{-\bx^2/a^2} \; ,
 \end{equation}
 \begin{equation}
   E_{gg}(\omega_n) = {1\over 2}\sum_{n=1}^{M-1} \omega_n
      - {1\over 2}\lambda M
         \left[{a^2d\over 2} + R_1^2\right]^{-d/2} \!\!\!
      + g^2 M \sum_{k=1}^{M} \left[{a^2d\over 2} + R_k^2\right]^{-d/2}\, .
 \end{equation} 
\end{enumerate}
The coefficient $\lambda$ in the third ($gg$) model is fixed by requiring 
that the bond length ($R_1$) for the bare chain ($g=0$) be the same 
in all three models,
\begin{equation}
  \lambda = {2a^2T_0^2\over d}\left(1 + {4\over\pi a^2T_0}\right)^{1+d/2} \; .
\end{equation}
The energies of  the three models were minimized numerically for various values
of $M$ and $g$. The results for $R^2$ and $R_1^2$ as functions of
the number of bits $M$ for various coupling constants $g$ are shown
in fig.~1.

\input psfig
\begin{figure}
\vskip -2.5cm
\centering
\hbox{\hskip-1.3cm\psfig{figure=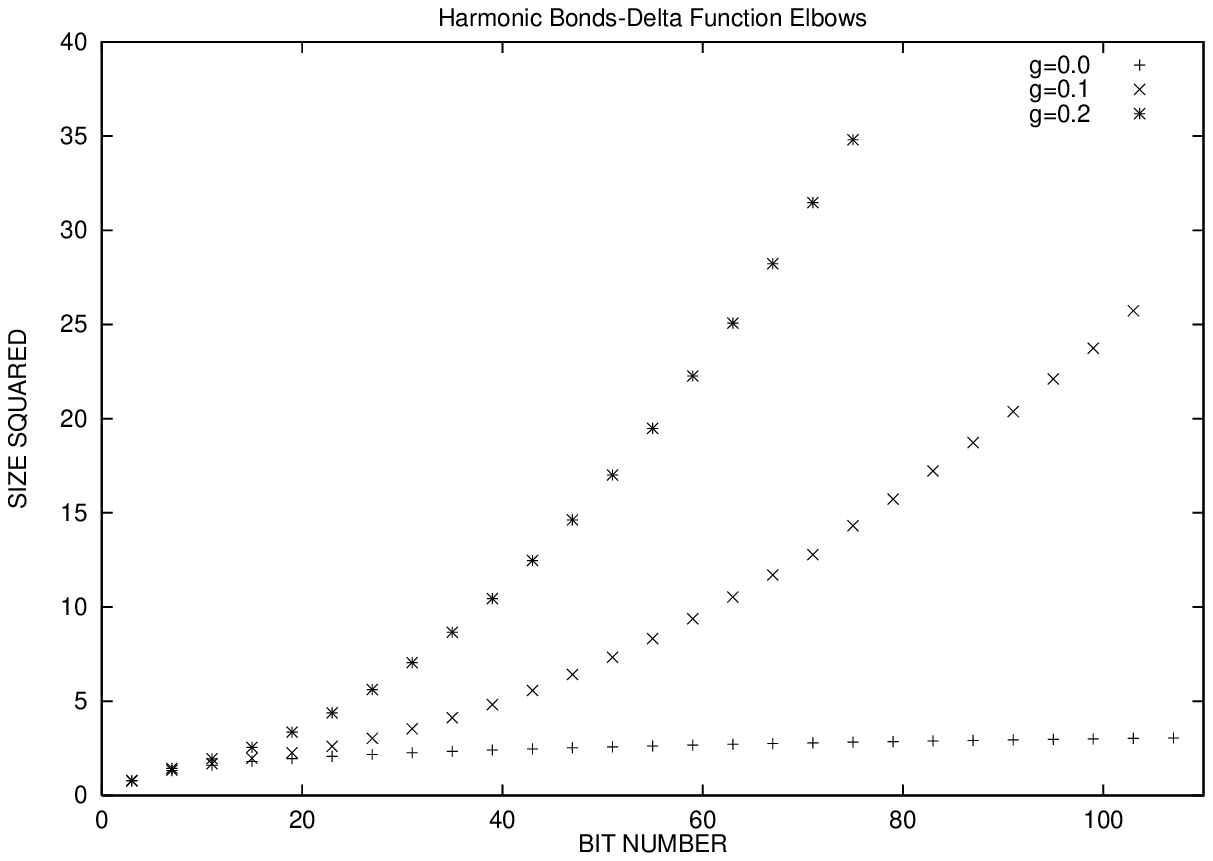,height=6.5cm}
\psfig{figure=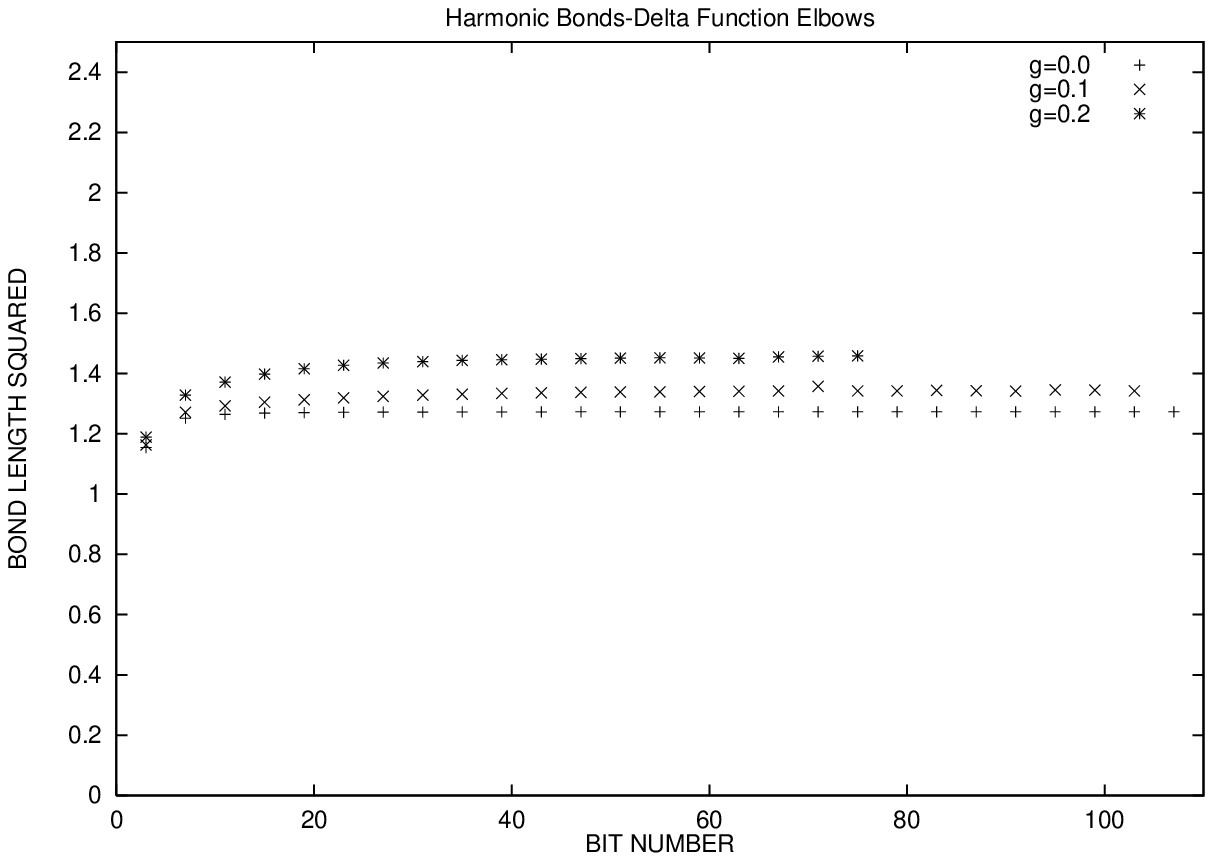,height=6.5cm}
}
\vskip 0.5cm
\hbox{\hskip-1.3cm\psfig{figure=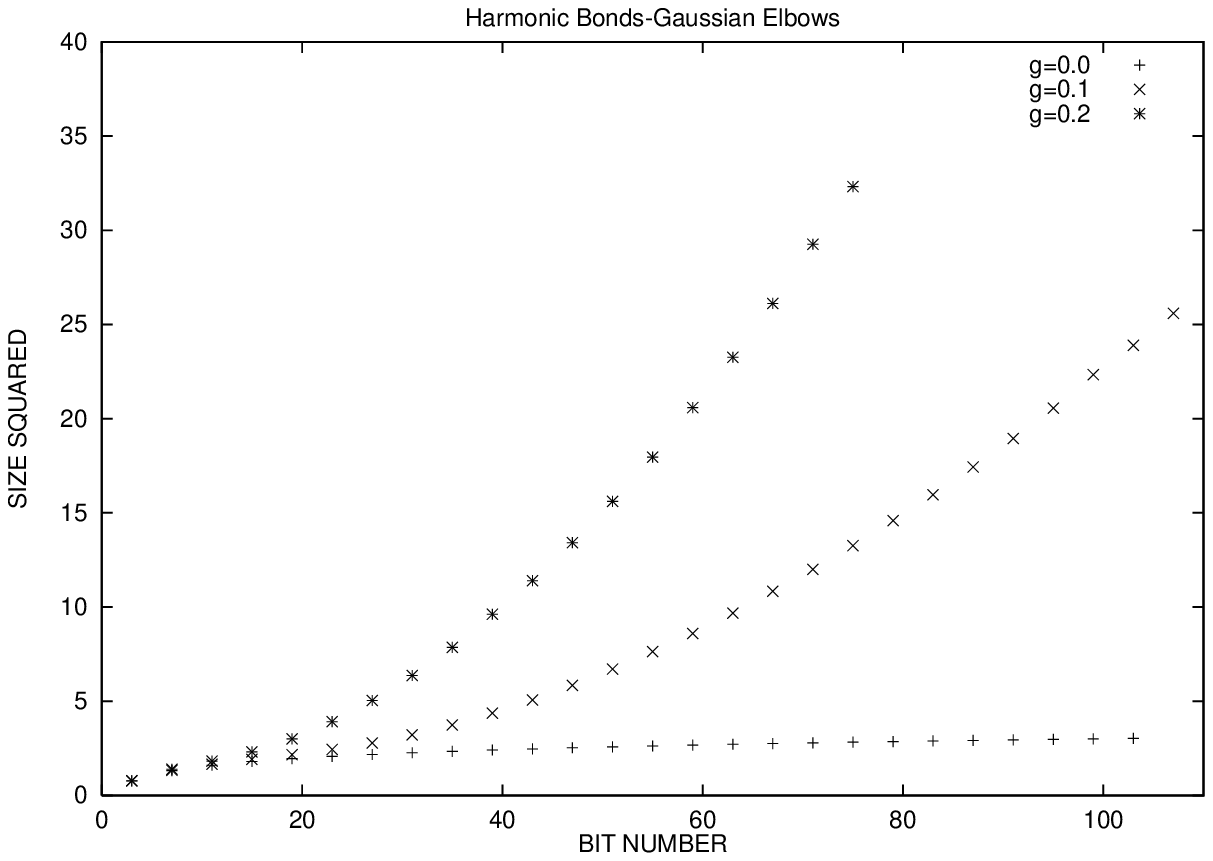,height=6.5cm}
\psfig{figure=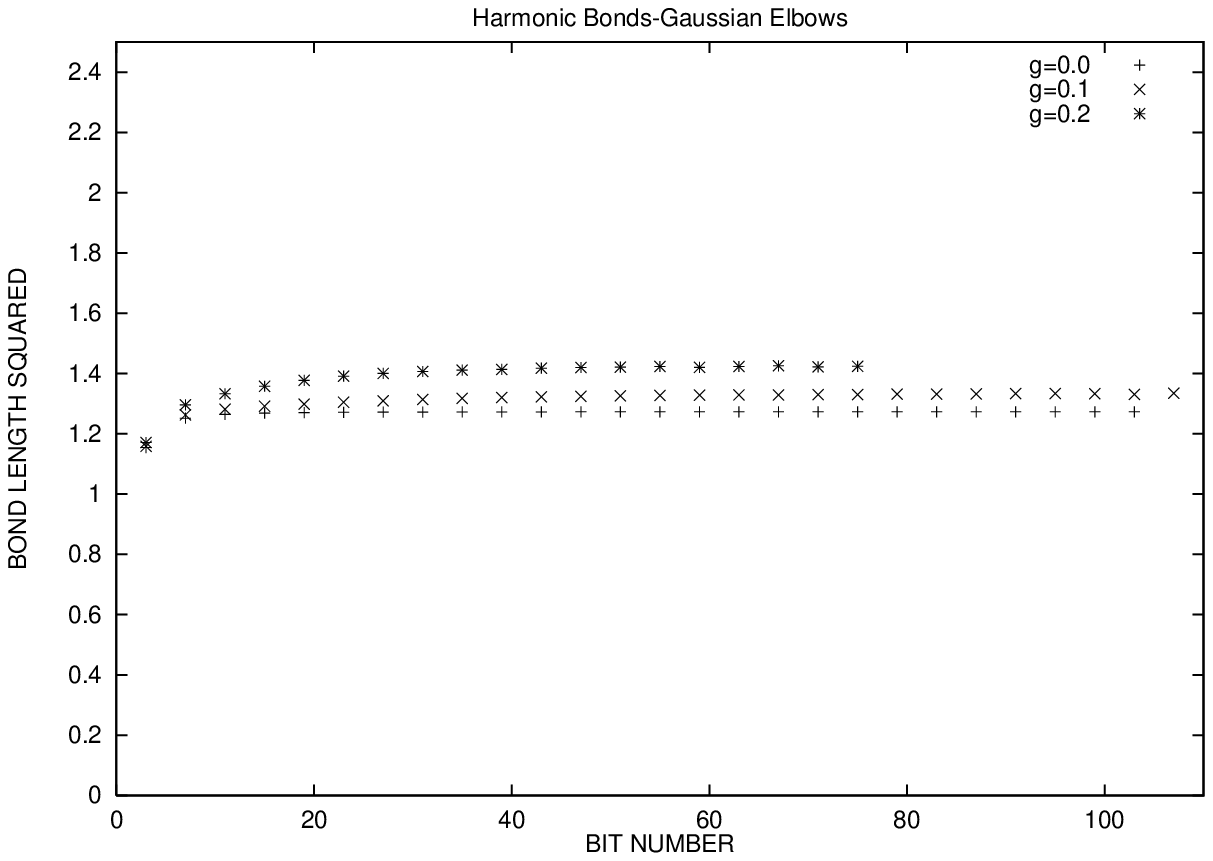,height=6.5cm}
}
\vskip 0.5cm
\hbox{\hskip-1.3cm\psfig{figure=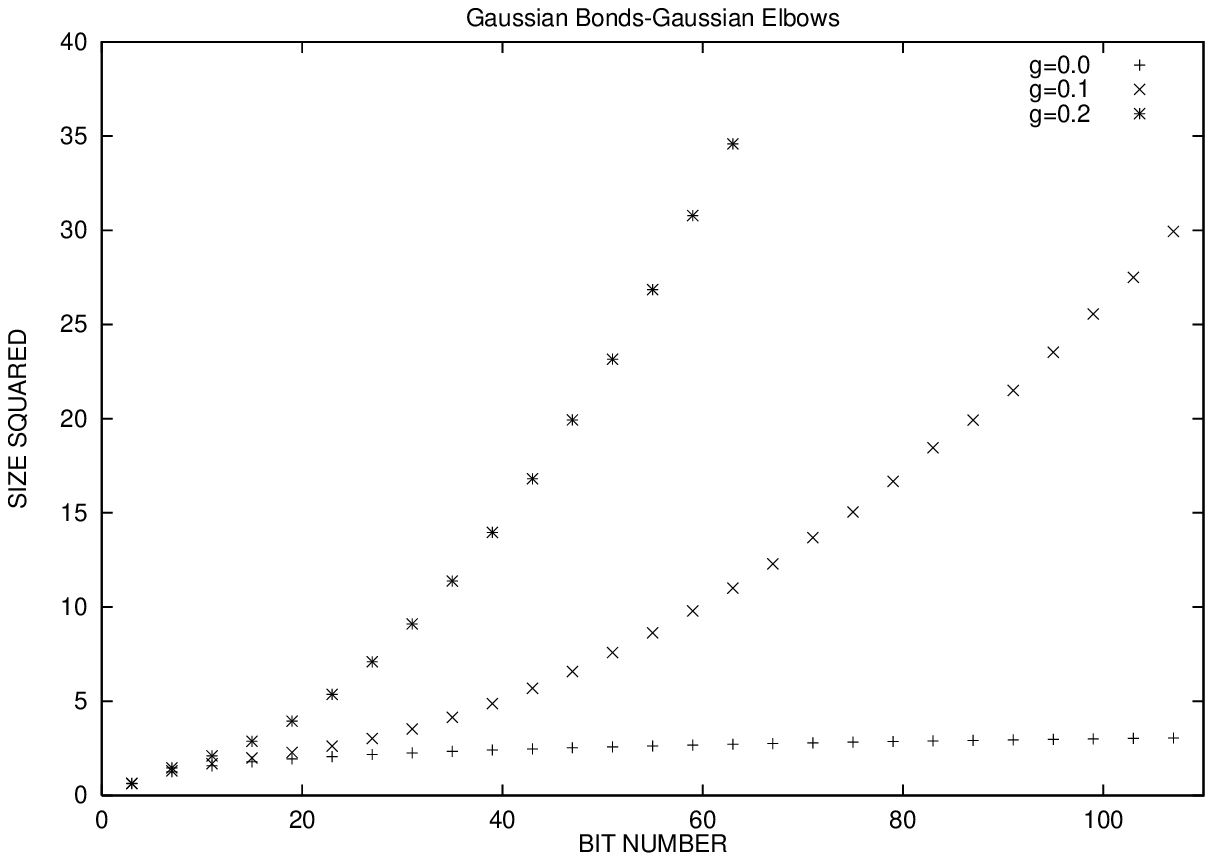,height=6.5cm}
\psfig{figure=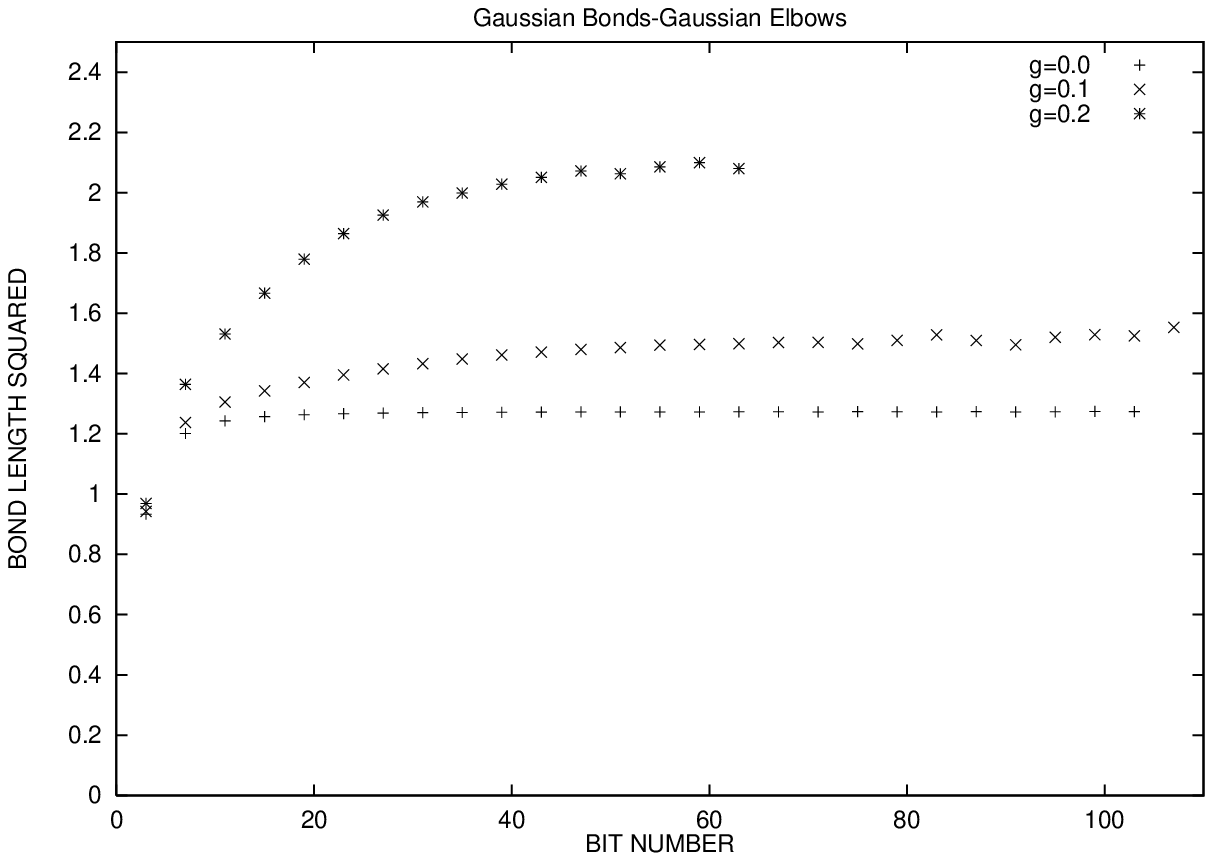,height=6.5cm}
}
\caption{Results of variational calculation of size and bond length
for various nearest-neighbor (bond) potentials and non-nearest-neighbor 
repulsions (elbows).}
\label{sizefig}
\end{figure}

The $R^2$ plots clearly exhibit a change in the 
growth pattern for $g\neq 0$ at some 
number of bits $M$. For small $M$ the growth is logarithmic,
in agreement with the fact that perturbation theory is valid there,
and the zeroth order result (\ref{baresize}) dominates. For large 
$M$ the growth is quadratic with $M$. 
This result seems to contradict the bound of linear growth
found in section 2.
Recall however that the linear bound was a consequence of the
gap scaling as $1/M$ at large $M$. From fig.~2 it is clear that 
for $g\neq 0$ the gap
in these models scales roughly as $1/M^3$ at large $M$. 
\begin{figure}
\vskip -2.5cm
\centering
\hbox{\hskip-1.3cm
\psfig{figure=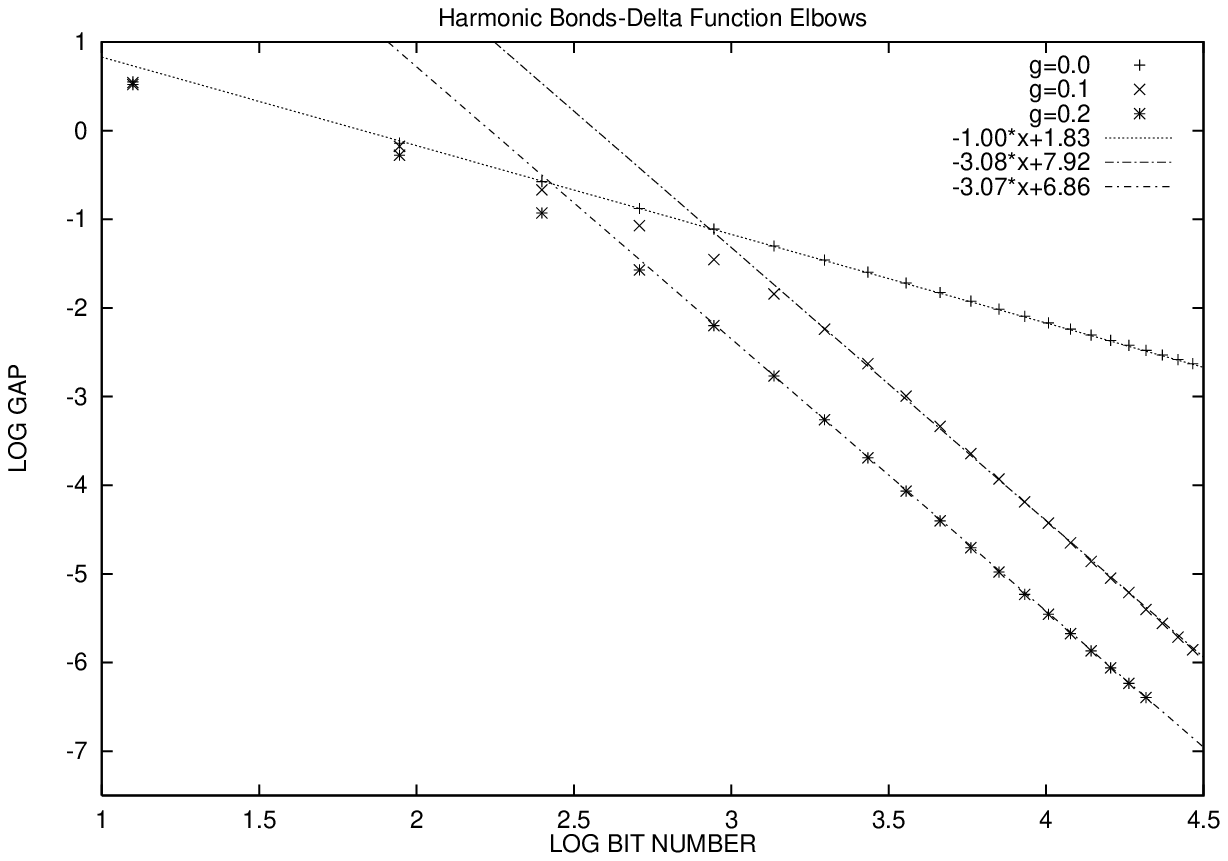,height=6.5cm}
\psfig{figure=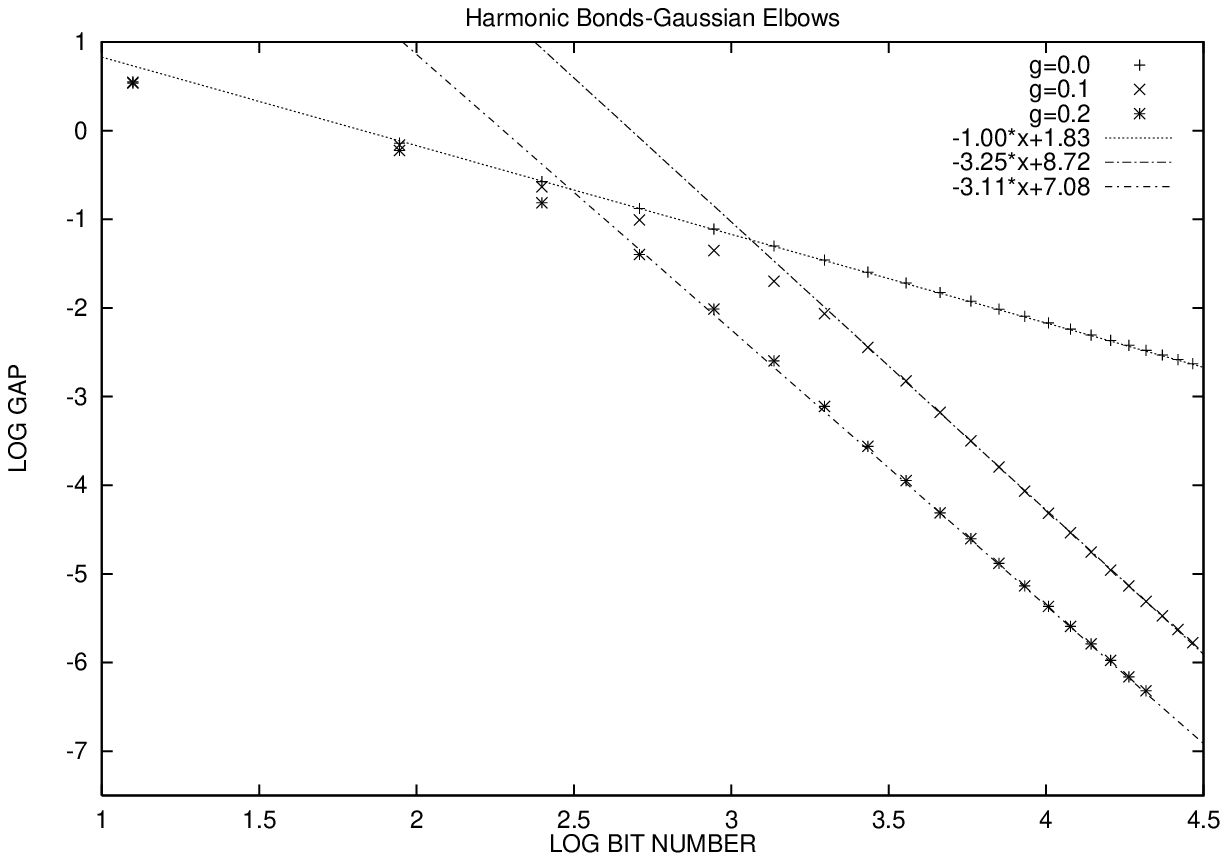,height=6.5cm}
}
\vskip 1cm
\hbox{%\hskip-1.3cm
\centerline{\psfig{figure=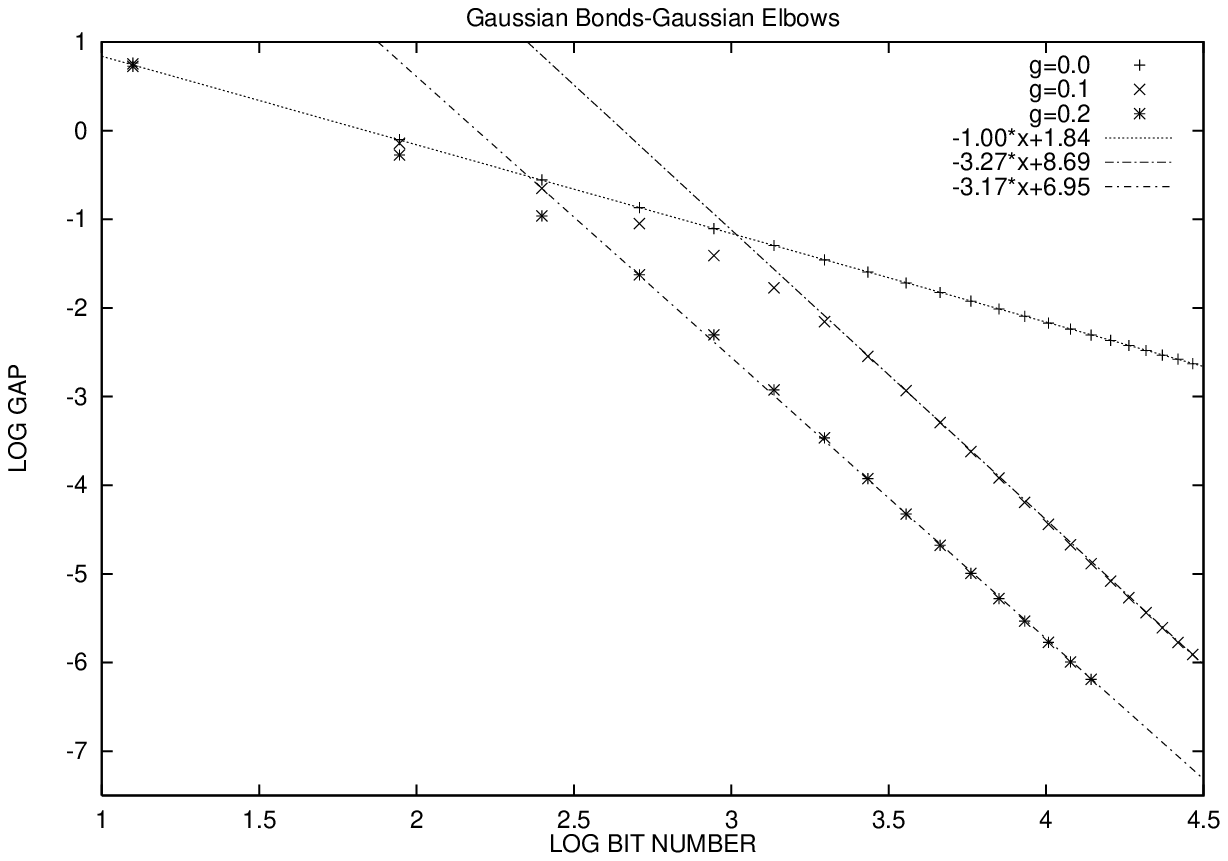,height=6.5cm}}
}
\caption{Gap dependence on bit number.}
\label{logfig}
\end{figure}
These polymers therefore do not correspond to relativistic strings 
in the continuum limit.

\section{Discussion}
\setcounter{equation}{0}
We have found that our analog string-bit model, simulating the
effects of string interactions on the size of a polymer of
string-bits, seems to predict a growth with bit
number that is much too fast for relativistic string.
However, there are several circumstances that indicate that the
analog model overstates the self-avoiding aspects of an
interacting string. First of all, the interactions of the
original bit model are much richer in structure than those
of the analog model. Our estimate of the net repulsive
character of the interactions (controlled in the analog
model by the value of $g$) involved 
the competition between huge numbers 
(of  order $M^2$) of repulsive and attractive contributions. 
A rough counting gave a net excess of repulsive
over attractive contributions of $O(1)$. For larger and
larger $M$ the repulsive and attractive contributions
are {\it relatively} more nearly in balance. 
Moreover, every contribution is not really
the same either quantitatively or qualitatively, so it is a bit
of a leap to conclude that the net result of the competition
is a pure repulsive interaction independent of $M$. It is conceivable that the
net repulsive interaction decreases with $M$.
This would correspond in our analog model to an effective $g$ depending in
some way on $M$. 

Furthermore, our analog model has vastly 
oversimplified the complicated bit rearrangement nature of the
interactions, replacing them with interactions that conserve
the integrity of each bit. Since a rearrangement term includes
a factor of the overlap of different wave functions, it is
bound to be smaller than a corresponding ``direct'' term. This
criticism of our analog model is made even sharper by the
circumstance that, in fact, {\it all} of the ``direct''
contributions exactly cancel out, leaving only contributions
which involve at least some bit rearrangement. Thus treating
the bit rearrangement properly would certainly reduce the growth
rate compared to our analog model. However, it is not 
at all clear whether the
reduction would amount to multiplying the size by a factor
smaller than unity but independent of $M$,
or by one that actually decreases with $M$. 

Finally, the models
studied in this paper are generic bosonic bit models. There is
no opportunity in these models for the further more subtle cancelations
that would be present in a supersymmetric model. We have tried
to include at least one implication of supersymmetry by 
starting with a manifestly {\it positive} Hamiltonian. Indeed
it was that positivity that was responsible for the net repulsive
character of the interactions in the first place. But imposing
positivity without supersymmetry is perhaps a bit heavy-handed, 
yet another reason to suspect that our analog model 
overstates the growth rate of string. Thus there is some hope that the
real string-bit model could predict a linear growth with bit number.

All of these issues need to be addressed. One could certainly try
to apply the variational method in the original bit model by varying
within the space of single bare polymer states. But handling bit
rearrangement in a tractable way remains a major challenge.
As long as bit rearrangement is taken into account, it is probably also
a good idea to try to include at least some admixture of
bare multi-polymer states in the trial wave function. Clearly,
this would require methods beyond those used in this article.
To study the role of supersymmetry, we must first
find a supersymmetric string-bit model. So far the only
candidate we have is the model in $d=1$ spatial dimension found
in \cite{bergmantoned}. 
In view of the discussion at the end of section 2
however, the $d=1$ case is not likely to shed light
on the size issue.
It is
probably most urgent to construct a supersymmetric bit
model in realistic dimensions ($d\geq2$).

\end{document}